\documentclass[prl,twocolumn,showpacs,amsmath]{revtex4-1}
\usepackage{hyperref}

\begin{document}

\title{Topological Kondo Effect in Transport through a Superconducting Wire\\
with Multiple Majorana End States}

\author{Oleksiy Kashuba}
\affiliation{Institute of Theoretical Physics, Technische Universit\"at Dresden,
01062 Dresden, Germany}
\author{Carsten Timm}
\affiliation{Institute of Theoretical Physics, Technische Universit\"at Dresden,
01062 Dresden, Germany}

\begin{abstract}
We investigate a system of multiple Majorana
states at the end of a topological superconducting wire coupled to a
normal lead. For a minimum of three Majorana fermions at the interface,
we find nontrivial renormalization physics. Interface tunneling processes can
be classified in terms of spin-$1/2$ and
spin-$3/2$ irreducible representations of the SU(2) group. We show
that the renormalization of the tunneling amplitudes belonging to different
representations is completely different in that one type is suppressed, whereas
the other is enhanced, depending on the sign of the Kondo-type
interaction coupling. This
results in distinct temperature dependencies of the tunneling current
through the interface and different
spin polarizations of this current.
\end{abstract}


\maketitle


\textit{Introduction}.---Majorana fermions were first proposed as
hypothetical elementary particles that are their own antiparticles
\cite{Majorana:1937}. The possibility of Majorana states at
the surfaces of triplet superconductors has been discussed for a long
time \cite{Lieb:1961,Kopnin:1991,Volovik:1999,Senthil:2000,Read:2000,
Beenakker:2013}.
The realization that they are related to topological properties of the
system \cite{Kitaev:2001} has generated a lot of interest in
Majorana states at the surfaces of topological
superconductors (TSs)
\cite{Schnyder:2008,Lutchyn:2010,Oreg:2010,Qi:2011,Alicea:2012,
Schnyder:2012,Schnyder:2013,Beenakker:2013}.

The first signatures of Majorana states at the ends of a
TS wire were found in transport measurements involving the interface
between the wire and a normal lead \cite{Mourik:2012,Deng:2012}.
%
%
%
These experiments have so far been compared to a model
with a single Majorana
state coupled to the normal lead \cite{Bolech:2007,Lutchyn:2010,Oreg:2010},
which cannot contain any interaction between the
Majorana state and the lead since the (single) Majorana operator $\gamma$
squares to unity. The theory has already advanced to more sophisticated
noninteracting systems, such as Josephson junctions between TSs, where the
tunneling takes place between Majorana states \cite{Sticlet:2013}, and
setups with one or several quantum dots mediating the electron
transfer between the leads and the TS \cite{Lu:2012,Cao:2012,Li:2013}.
The study of interaction processes in such systems is of interest since
interactions generically lead to strong renormalizations in low dimensions.
However, so far only on-dot interactions have been studied for these
setups \cite{Golub:2011,Lee:2013}.
The implementation of Majorana-lead interactions requires the presence
of several Majorana modes. Multiple Majorana states and the
renormalization of interaction couplings have been studied in
Refs.\
\cite{Beri:2012,Zazunov:2013,Altland:2014,Eriksson:2014}. Each
Majorana end state is either coupled by a tunneling term to its own
normal lead or is not coupled at all
\cite{Beri:2012,Zazunov:2013,Altland:2014,Eriksson:2014}. We consider
a different situation: multiple Majorana states hybridizing
with a single lead.

Our goal is to understand the interplay between multiple
tunneling channels and the electron-Majorana interaction,
which we find to induce strong renormalization.
This research is meant to help in interpreting, regardless of
microscopic details, the results of transport
measurements by studying the temperature dependence and spin
polarization of the current. We show that these observables exhibit clear
signatures of the presence of Majorana fermions and of their coupling to the
leads. A TS wire coupled to a normal lead is modeled by
$N$ Majorana fermions localized at one end of the wire
and a Fermi sea of spinful electrons,
coupled by general tunneling and interaction terms.
The minimal nontrivial case of $N=2$ gives nothing new since
the two Majorana states make up a spinless fermion and the
interaction in the system is equivalent to the one in the
interacting resonant-level model, leading to the same
renormalization flow, which has been studied extensively
\cite{Tsvelick:1983,*Schlottmann:1982a,*Schlottmann:1982b,*Schlottmann:1982c,
Andergassen:2011,*Kashuba:2013}.
Systems with $N\geq3$ are fundamentally different: unlike $N=2$
system, their interaction couplings get strongly renormalized, similarly
to the Kondo model \cite{Eriksson:2014}.
Here, we will demonstrate that interesting renormalization physics occurs
already for $N=3$.
The predictions made in this work are unique for this system, which
supports both Kondo and tunneling couplings, whose interplay
leads to the non-trivial discrimination of the tunneling processes
depending on the sign of the interaction.

In the general case of $N$ Majorana states, the sets of
$N$ Majorana operators before and after some symmetry transformation are
related by $\gamma_{i}'=\sum_{j=1}^{N}R_{ij}\gamma_{j}$,
where $R$ is a real (since $\gamma^{\dagger}_{i}=\gamma_{i}$)
orthogonal matrix belonging to the group SO($N$).
A candidate for this symmetry transformation
is the electron spin rotation.
In this case Majorana states transform into each other according
to a representation of the SU(2) group, which also has to be
a subgroup of SO($N$).
The case of three Majorana states is particularly interesting since
the whole SO(3) group is equivalent to the spin-1 representation of
SU(2). An experimental realization of a set of three Majorana states
transforming
under SO(3) is still unknown, but there is already a proposal assuming
the existence of such sets in vortex cores in TS
\cite{Yasui:2011}. As we shall see, for the N--TS
interface the tunneling terms inevitably break the SU(2) spin symmetry.

In this paper, we derive the renormalization-group (RG) flow equations
for the electron-Majorana interaction strengths and tunneling amplitudes
within the framework of poor man's scaling for arbitrary $N$.
We solve the RG equations for the simplest nontrivial case $N=3$ and
demonstrate that the tunneling amplitudes can be classified according to
the irreducible representations of the SU(2) group and that the components
belonging to different representations obey different RG equations.
In practice, this means that starting from arbitrary tunneling parameters,
the interaction will lead to the suppression of one set of parameters and the
enhancement of the other.
Moreover, depending on the initial value of the interaction,
a different tunneling type will dominate in the scaling limit, leading
to a different temperature dependence of the current through the interface.



\textit{Model}.---The investigated system consists of a
noninteracting normal lead with a Fermi sea of electrons coupled
to three Majorana states localized at the same end of a TS wire, which
are described by the Hermitian fermionic operators $\gamma_{i}$.
The Hamiltonian of the lead is
$H_{L}= \sum_{\alpha\mathbf{p}} \epsilon_{\mathbf{p}}\,
  a^{\dagger}_{\alpha\mathbf{p}}a_{\alpha\mathbf{p}}$,
where $a^\dagger_{\alpha\mathbf{p}}$, $a_{\alpha\mathbf{p}}$ are creation
and annihilation operators of electrons with spin
$\alpha={\uparrow}$, ${\downarrow}$ and momentum $\mathbf{p}$.
It is assumed that
the electronic band with the dispersion relation $\epsilon_{\mathbf{p}}$
approximately covers the energy interval $[-D,\,D]$ and has a
constant normalized density of states
$\rho(E)\equiv\mathcal{N}^{-1}\sum_{\mathbf{p}}
\delta(E-\epsilon_{\mathbf{p}})\approx\nu$ for $E\ll D$
(here $\mathcal{N}$ is a total number of states in the lead).
Henceforth, we take $\hbar=k_B=1$.

The couplings between the states localized at opposite ends of the wire
are exponentially suppressed with the distance between them.
If the SO($N$) symmetry of the Majorana states $\gamma_{i}$
at the same end
is broken, a coupling of the form
$H_{D}=i\sum_{ij}E_{ij}\gamma_{i}\gamma_{j}$ is allowed. However, as we will
discuss later, $H_D$ does not affect the RG equations as long
as the flow parameter satisfies $\Lambda\gg |E_{ij}|$.

The N--TS coupling consists of a bilinear tunneling part and an
interaction part. Assuming that the coupling is local in real
space, the most general tunneling term is
\begin{equation}
H_{T} = \sum_{i\alpha} t_{i\alpha}\, \gamma_{i}\, a_{\alpha}^{\dagger}
  + \text{H.c.},
\label{eq:Htunneling}
\end{equation}
where $a_{\alpha}=\mathcal{N}^{-1/2}\sum_{\mathbf{p}}a_{\alpha\mathbf{p}}$
and the $t_{i\alpha}$ are tunneling amplitudes.
The leading interaction terms are of fourth order in fermionic operators.
We focus on terms that are quadratic in Majorana operators~\footnote{In
principle, there are also N--TS interaction terms of the form 
$\gamma_{i} a_{\uparrow}a_{\downarrow}a_{\alpha}^{\dagger}+\mathrm{H.c.}$ and
$\gamma_{i}\gamma_{j}\gamma_{k} a_{\alpha}^{\dagger}+\mathrm{H.c.}$,
which can be interpreted as interaction corrections to tunneling, and
$\gamma_{i}\gamma_{j}a_{\uparrow}a_{\downarrow}+\mathrm{H.c.}$,
which is an Andreev tunneling term. These more exotic terms can be treated in
the same manner.
We find that these terms are generally renormalized and produce
corrections to the tunneling and the biquadratic interaction,
but we assume these couplings to be small for simplicity.}.
Due to the anticommutation relation $\{\gamma_{i},\gamma_{j}\}=2\delta_{ij}$
only $N(N-1)/2$ combinations exist.
Thus, the most general local biquadratic interaction term reads
\begin{equation}
H_{V}= \frac{1}{2}\sum_{ij\alpha\beta} V^{ij}_{\alpha\beta}\,
  \gamma_{i}\gamma_{j} a^{\dagger}_{\alpha}a_{\beta},
\label{eq:Hinteraction}
\end{equation}
where $V^{ij}_{\alpha\beta}=-V^{ji}_{\alpha\beta}$ are coupling
parameters.
If there is any interaction between the TS and the leads,
we expect an expansion in the order of vertices to generate $H_V$. While the
direct Coulomb interaction vanishes for the neutral Majorana fermions,
an exchange-type interaction emerges naturally since the
zero-energy
Majorana surface states of nodal TSs with strong spin-orbit coupling typically
carry a large spin \cite{Brydon:2013,Schnyder:2013,Brydon:2014}.
An interaction $H_V$ can also be realized in a small
superconducting island with large charging energy hybridized with normal
leads \cite{Beri:2012,Altland:2014}. $H_V$ is here obtained by integrating out
charge fluctuations, which removes the tunneling term $H_T$. More generally,
the coupling of Marojana states and normal electrons to any additional
modes, such as phonons, will typically introduce an effective
interaction of this form when these modes are integrated out.





\textit{RG and symmetry analysis}.---To study the
renormalization effects, we employ the poor man's
scaling approach \cite{Wilson:1975,Borda:2007}:
the RG flow parameter $\Lambda$ denotes the maximal energy of the electron
modes, $|\epsilon_{\mathbf{p}}|<\Lambda$; the electron modes are
divided into \emph{fast} modes $a_{\alpha\mathbf{k}}$ with energies
in the thin shell $\Lambda-\Delta\Lambda < |\epsilon_{\mathbf{k}}| < \Lambda$
and \emph{slow} modes $a_{\alpha\mathbf{p}'}$ with
$|\epsilon_{\mathbf{p}'}|<\Lambda-\Delta\Lambda$;
integration over the fast modes results in corrections to the slow-mode
terms in the Hamiltonian. Repeating this step, we integrate
out all electron degrees of freedom, obtaining an effective low-energy
Hamiltonian. Taking the N--TS coupling as the
perturbation and $H_{0}=H_{L}+H_{D}$ as the bare Hamiltonian,
the correction to the interaction for excitations with small
energy $E$, from a single RG step, reads
\begin{eqnarray}
\Delta H_{V} &\approx& \langle
  H_{V}(E-H_{0})^{-1}H_{V}\rangle
 = - \frac{1}{4\mathcal{N}^{2}} \!\!\!\! \sum_{ii'jj',\alpha\beta\eta \atop \mathbf{p'q';k}} \!\!
 V_{\alpha\eta}^{ij}V_{\eta\beta}^{i'j'} \nonumber \\
&\times& \!\left(\!
 \gamma_{i}\gamma_{j}\gamma_{i'}\gamma_{j'}  \frac{1-n_{\mathbf{k}}}{\epsilon_{\mathbf{k}}}
  + \gamma_{i'}\gamma_{j'}\gamma_{i}\gamma_{j} \frac{n_{\mathbf{k}}}{\epsilon_{\mathbf{k}}}
 \!\right)\!
 a^{\dagger}_{\alpha \mathbf{p}'}a_{\beta \mathbf{q}'},\qquad
\label{eq:rg.DeltaHV}
\end{eqnarray}
where $\mathbf{p}'$, $\mathbf{q}'$ denote slow modes, $\mathbf{k}$
refers to a fast mode, angular brackets denote the integration over the
fast modes only, $\Delta H \equiv H(\Lambda-\Delta\Lambda)-H(\Lambda)$ is
the difference between the values after and before the RG step,
and $n_{\mathbf{k}}\equiv n_{F}(\epsilon_{\mathbf{k}})$ is a Fermi
distribution function.
$\Lambda$ is of the order of the band width, which is
assumed to be large compared to the other energy scales of the problem, in
particular the energy $E$ and the inter-Majorana couplings $E_{ij}$. Therefore,
these terms do not affect the RG flow to leading order and can be neglected.
The terms relevant for the RG flow decay as
$\Lambda^{-1}$. For the assumed constant
and symmetric density of states we drop the sum
$\mathcal{N}^{-1} \sum_{\mathbf{\mathbf{k}}}1/\epsilon_{\mathbf{\mathbf{k}}}$ and
approximate
$\mathcal{N}^{-1} \sum_{\mathbf{k}}(1/2-n_{\mathbf{k}})/\epsilon_{\mathbf{k}}\approx \nu\,
\Delta\Lambda/\Lambda$, and find the RG equation
\begin{equation}
\frac{d V_{\alpha\beta}^{ij}}{d\Lambda} = \frac{2\nu}{\Lambda}
\sum_{l,\eta}\left(V_{\alpha\eta}^{il}V_{\eta\beta}^{lj}
  - V_{\alpha\eta}^{jl}V_{\eta\beta}^{li}\right) .
\label{eq:rgsV}
\end{equation}
The corresponding correction to the tunneling term is
\begin{eqnarray}
\lefteqn{ \Delta H_{T} \!\approx\!\langle
  H_{V}(E \!-\! H_{0})^{-1}H_{T}\rangle \!+\! [T\leftrightarrow V] \!=\! - \frac{1}{2\mathcal{N}^{3/2}} }
  \nonumber \\
&& \!\!\times\!\!\! \sum_{ijj',\eta\beta \atop \mathbf{p}';\mathbf{k}} \!\!
  V_{\alpha\beta}^{ij} t_{j'\beta} \!\left(\!
   \gamma_{i}\gamma_{j}\gamma_{j'} \frac{1 \!-\! n_{\mathbf{k}}}{\epsilon_{\mathbf{k}}}
  \!+\! \gamma_{j'}\gamma_{i}\gamma_{j} \frac{n_{\mathbf{k}}}{\epsilon_{\mathbf{k}}}
  \!\right)\! a_{\alpha\mathbf{p}'}^{\dagger} \!+\! \text{H.c.} \quad
\label{eq:rg.DeltaHT}
\end{eqnarray}
Keeping only the RG-relevant contribution, we obtain
\begin{equation}
\frac{d t_{i\alpha}}{d\Lambda} = \frac{2\nu}{\Lambda} \sum_{j,\beta} V_{\alpha\beta}^{ij}t_{j\beta}.
\label{eq:rgsT}
\end{equation}
The obtained equations couple $2N(N-1)$ quantities
$V^{ij}_{\alpha\beta}$ and $2N$ quantities $t_{i\alpha}$.
To simplify the analysis but preserve the interesting renormalization
physics, we restrict ourselves to $N=3$.



\textit{The case of three Majorana states}.---The special feature
of the SO(3) group is
that its irreducible representations are equivalent to integer-spin
representations of SU(2).
This feature allows us to classify the elements $V^{ij}_{\alpha\beta}$
and $t_{i\alpha}$ in terms of
the irreducible representations of SU(2).
The products of two Majorana operators, which form vectors
belonging to the spin-1 representation $\Gamma_{1}$, can be split into the
irreducible representations
$\Gamma_{1}\otimes\Gamma_{1}\cong\Gamma_{0}\oplus\Gamma_{1}\oplus\Gamma_{2}$.
Since expressions belonging to the scalar ($\Gamma_{0}$) representation,
$\sum_{i}\gamma_{i}^{2}=3$, and to the spin-2 ($\Gamma_{2}$)
representation, $\gamma_{i}\gamma_{j}+\gamma_{j}\gamma_{i}=0$, are just numbers,
the only non-trivial combination is the Majorana pseudospin
operator $s^{M}_{i}=-(i/2)\sum_{jj'}\epsilon_{ijj'}\gamma_{j}\gamma_{j'}$
(here $\epsilon_{ijj'}$ is the three-dimensional Levi-Civita tensor),
which belongs to the $\Gamma_{1}$ representation of SU(2).
The operators $s^M_i$ play the role of pseudospin components; they
satisfy the algebra
$[s_{j}^{M},s_{j'}^{M}]=2i\sum_i\epsilon_{ijj'}s_{i}^{M}$ and
$[s^{M}_{j},\gamma_{j'}]=2i\sum_i\epsilon_{ijj'}\gamma_{i}$.
Expressed in these terms, the interaction term in
Eq.~\eqref{eq:Hinteraction} takes the form
\begin{equation}
H_{V}=\sum_{i}M_{i}\, s^{M}_{i} n^{L} + \sum_{ij}V_{ij}\, s^{M}_{i}s^{L}_{j} ,
\label{eq:Hintsplit}
\end{equation}
where $n^{L}=\mathcal{N}^{-1}\sum_{\alpha,\mathbf{pq}}
a_{\alpha\mathbf{p}}^{\dagger} a_{\alpha\mathbf{q}}$ is the local
lead-electron number operator
and $s^{L}_{i}=\mathcal{N}^{-1}\sum_{\alpha\beta,\mathbf{pq}}
a_{\alpha\mathbf{p}}^{\dagger}\sigma_{\alpha\beta}^{i}a_{\beta\mathbf{q}}/2$ the
corresponding spin operator, where $\sigma^{i}$ are Pauli matrices.
The first term, when substituted into Eq.~\eqref{eq:rgsV}, is not
renormalized and just leads to a renormalization of the
tunneling amplitudes through Eq.~\eqref{eq:rgsT}, similar to the
interacting resonant-level model \cite{Andergassen:2011}.
Setting the vector $M_{i}$ to $(0,0,M_{z})$ by choosing an appropriate
basis, we find that the $z$-component of the tunneling amplitude does
not change, $t_{\alpha z}(\Lambda)=t_{\alpha z}$, while the others are
renormalized as $t_{\alpha,\pm}(\Lambda)=t_{\alpha,\pm}(D/\Lambda)^{\pm
2\nu M_{z}}$, where $t_{\alpha,\pm}=t_{\alpha x}\pm i\, t_{\alpha y}$
\cite{Zazunov:2013}.
The second term in Eq.~\eqref{eq:Hintsplit} contains the product of two
vectors, so it can be decomposed as $V_{ij} =
\delta_{ij} J+ \sum_{k} \epsilon_{ijk} J^{k} + J^{ij}$, where $J$ is
a scalar ($\Gamma_{0}$), which describes the Kondo-type interaction between the lead
electrons and the effective Majorana spin, $J^{i}$ is a vector ($\Gamma_{1}$), and
the symmetric matrix $J^{ij}$ with zero trace corresponds to the spin-2
representation $\Gamma_{2}$.

Since the main goal of this paper is to demonstrate the possibility of
interesting renormalization physics, we restrict
ourselves to the simplest case with unbroken SU(2) symmetry in the interaction
between normal lead and TS, choosing $V_{ij}=\delta_{ij}J$.
Then Eq.~\eqref{eq:rgsV} leads to the well-known RG flow equation for
the Kondo coupling \cite{Wilson:1975,Beri:2012,Zazunov:2013},
\begin{equation}
\frac{dJ}{d\Lambda}=-\frac{2\nu J^{2}}{\Lambda}.
\label{eq:rgJ}
\end{equation}
The solution depends on the sign of the initial
unrenormalized coupling $J_{0}$ (we denote initial values by a
subscript $0$): The
coupling is enhanced for $J_{0}>0$ and suppressed for $J_{0}<0$,
depending on $\Lambda$ as
\begin{equation}
J = \frac{1}{2\nu\, \ln(\Lambda/T_{K})},
\label{eq:rgJsolution}
\end{equation}
where $T_{K}=D\, e^{-1/2\nu J_{0}}$ is the Kondo temperature.
The poor man's scaling approach, however, breaks down when $\Lambda$ reaches
the largest of the low-energy scales of the problem, $\Lambda_{c}$,
which plays the role of an infrared cutoff.
For the antiferromagnetic case ($J>0$)
this means that $J$ actually saturates and does not
diverge at $\Lambda=T_{K}$, as Eq.~\eqref{eq:rgJsolution} would predict
\cite{Andergassen:2011,Pletyukhov:2012,Wilson:1975}.
In the context of a possible implementation utilizing spin-polarized
Majorana surface states, it is plausible that either sign of $J$ can be
realized since model calculations find Majorana states in pairs with opposite
spin expectation value \cite{Schnyder:2013,Brydon:2014}.

The tunneling term in Eq.~\eqref{eq:Htunneling} contains a product of a
vector and a spinor. Thus the
tunneling amplitudes can be classified by the irreducible
representations of SU(2), $\Gamma_{1} \otimes \Gamma_{1/2} \cong
\Gamma_{1/2} \oplus \Gamma_{3/2}$, and split
into spin-1/2 and spin-3/2 terms according to
$t_{i\alpha}=\sum_{S,m}t^{S,m} \tau_{i\alpha}^{S,m}$, where
$m=\pm1/2$ for $S=1/2$ and
$m=\pm1/2,\pm3/2$ for $S=3/2$. The Clebsch-Gordon coefficients  for
$S=1/2$ read
\begin{equation}
\tau^{\frac{1}{2},+\frac{1}{2}}_{i\alpha} \!=\! \tfrac{1}{\sqrt{3}}\!
\begin{pmatrix}
 0 & 0 & 1 \\
 1 & i & 0
\end{pmatrix}_{\!\!\alpha i} \!, \,
\tau^{\frac{1}{2},-\frac{1}{2}}_{i\alpha} \!=\! \tfrac{1}{\sqrt{3}}\!
\begin{pmatrix}
 1 & -i & 0 \\
 0 & 0 & -1
\end{pmatrix}_{\!\!\alpha i} \!,
\label{eq:matr12}
\end{equation}
which are basically Pauli matrices with swapped indices,
$\tau_{i\alpha}^{1/2,+1/2}=\sigma^{i}_{\alpha,\uparrow}$,
$\tau_{i\alpha}^{1/2,-1/2}=\sigma^{i}_{\alpha,\downarrow}$. The
Clebsch-Gordon coefficients for $S=3/2$ are
\begin{equation}
\begin{split}
&
\tau^{\frac{3}{2},+\frac{3}{2}}_{i\alpha} =
\begin{pmatrix}
 \frac{1}{2} & \frac{i}{2} & 0 \\
 0 & 0 & 0
\end{pmatrix}_{\!\!\alpha i}\!, \,
\tau^{\frac{3}{2},+\frac{1}{2}}_{i\alpha} =
\tfrac{1}{\sqrt{3}}
\begin{pmatrix}
 0 & 0 & -1 \\
 \frac{1}{2} & \frac{i}{2} & 0
\end{pmatrix}_{\!\!\alpha i}\!,
\\
&
\tau^{\frac{3}{2},-\frac{1}{2}}_{i\alpha} =
\tfrac{1}{\sqrt{3}}
\begin{pmatrix}
 -\frac{1}{2} & \frac{i}{2} & 0 \\
 0 & 0 & -1
\end{pmatrix}_{\!\!\alpha i}\!, \,
\tau^{\frac{3}{2},-\frac{3}{2}}_{i\alpha} =
\begin{pmatrix}
 0 & 0 & 0 \\
 -\frac{1}{2} & \frac{i}{2} & 0
\end{pmatrix}_{\!\!\alpha i} .
\end{split}
\label{eq:matr32}
\end{equation}
The absence of the scalar representation for the tunneling amplitudes
between spin-1/2 electrons and triplets of Majorana states
signifies that the SU(2) is always broken,
as mentioned in the introduction.

According to Eq.~\eqref{eq:rgsT}, the tunneling coefficients obey
the RG equations
\begin{equation}
\frac{dt^{1/2,m}}{d\Lambda} = -\frac{4\nu J t^{1/2,m}}{\Lambda} , \quad
\frac{dt^{3/2,m}}{d\Lambda} = \frac{2\nu J t^{3/2,m}}
  {\Lambda}.
\label{eq:rgt}
\end{equation}
Together with Eq.~\eqref{eq:rgJsolution}, the solutions read
\begin{equation}
t^{1/2,m} = t^{1/2,m}_0\, \frac{J^2}{J_0^2} , \quad
t^{3/2,m} = t^{3/2,m}_0\, \frac{J_0}{J} .
\label{eq:rgtsolution}
\end{equation}
For antiferromagnetic coupling, spin-3/2 tunneling is suppressed, whereas
spin-1/2 tunneling rapidly increases as $\Lambda$ approaches $T_K$,
together with the Kondo coupling $J$.
Ferromagnetic coupling leads to the opposite behavior:
spin-3/2 tunneling increases, while spin-1/2 tunneling decreases.
The physical values of the renormalized parameters are obtained at
the end of the RG flow.
Although Eqs.~\eqref{eq:rgJsolution} and \eqref{eq:rgtsolution}
generally break down at the infrared cutoff $\Lambda_{c}$,
if the temperature $T$ is much larger than all other low-energy
scales (but still much smaller than ultraviolet cutoff $D$),
the renormalized coupling parameters can be obtained by
substituting the flow parameter by the temperature, $\Lambda=T$.


\textit{Results and discussion}.---The results in
Eq.~\eqref{eq:rgtsolution} demonstrate that antiferromagnetic
coupling at the interface enhances the transport
with smaller total spin, while ferromagnetic coupling enhances
the tunneling transport with larger total spin. In the general case
when the initial Hamiltonian contains all possible tunneling amplitudes,
the presence of a Kondo interaction leads to a strong renormalization,
which manifests itself by an
instability of the tunneling amplitudes. Independently
of the coupling sign, the total tunneling probability is enhanced.
However, if the
initial interaction is antiferromagnetic the system is dominated by spin-1/2
tunneling, while for ferromagnetic interaction it is dominated by spin-3/2
tunneling.
The type of coupling thus manifests itself in transport
processes. One of its signatures is the temperature dependence of the
current through the N--TS interface. For voltages
$U$ much larger than the temperature but smaller than the
superconducting gap, the current $I$ is proportional to the tunneling
probability, $I\propto |t|^{2}$ \cite{Bolech:2007}.
According to Eqs.~\eqref{eq:rgJsolution}
and \eqref{eq:rgtsolution}, the current thus depends on
temperature as $I\propto \ln^{-4} (T/T_{K})$ for the
antiferromagnetic case and as $I\propto \ln^2 (T_{K}/T)$ for the
ferromagnetic case. This provides us with a criterion for the
detection of multiple Majorana states and for determining the
type of interaction between normal lead and TS.

The dominant renormalized spin-$S$ tunneling also leads to a distinctive
spin dependence of the current through the interface.
For spin-1/2 tunneling (antiferromagnetic case), two of the three Majorana
fermions can be combined into one conventional (Dirac) fermion
$d=\frac{1}{2}(\gamma_{x}+i\gamma_{y})$ so that the tunneling
Hamiltonian becomes
\begin{eqnarray}
H_{T} &=& \sum_{i\alpha m} t^{1/2,m} \tau^{1/2,m}_{i\alpha} \gamma_i\,
a^\dagger_\alpha + \mathrm{H.c.} \nonumber \\
&=& t'_1\, (- \gamma_z a^\dagger_\downarrow
  + 2d^\dagger a^\dagger_\uparrow)
  + t'_2\, ( \gamma_z a^\dagger_\uparrow
  + 2d a^\dagger_\downarrow ) + \mathrm{H.c.} ,\qquad
\label{eq:Hd12}
\end{eqnarray}
where $t'_1 \equiv t^{1/2,-1/2}/\sqrt{3}$, $t'_2 \equiv
t^{1/2,1/2}/\sqrt{3}$.
The tunneling amplitudes $t^{1/2,\pm 1/2}$ form a spinor, so their
component values depend on the choice of basis in spin space.
By an appropriate choice one can always set one of the elements
$t_{n}'$ to zero.
Upon setting $t_{1}'=0$, the system decomposes into two noninteracting
parts. The first one consists of spin-up electrons bound to the $\gamma_z$
Majorana state, while the second is a resonant-level model made up of
spin-down electrons and the additional fermion $d$.
We now discuss the contributions of the two parts to the tunneling current
under a bias voltage. The first part allows a nonzero stationary current, as we
can see as follows: The Majorana operator can be expressed in terms of Dirac
operators as $\gamma_z = d'+(d')^\dagger$. Thus the combined particle number
$a^\dagger_\uparrow a_\uparrow + (d')^\dagger d'$ is not conserved. If we
assume, to be specific, a positive bias voltage to be applied to the TS,
spin-up electrons will tunnel into the TS alternatingly
creating and annihilating the $d'$ fermion. Physically, this represents Andreev
tunneling~\cite{Bolech:2007}; the charge conservation is
restored by the creation of Cooper pairs in the superconducting condensate. On
the other hand,
the second part of the model does conserve the combined particle number
$a^\dagger_\downarrow a_\downarrow + d^\dagger d$ and the $d$ fermion is not
connected to any other lead. Thus the stationary current for the spin-down
electrons vanishes. In conclusion,
the spin-1/2 coupling results in a fully spin-polarized current in
the basis defined by the tunneling-amplitude spinor.

For spin-3/2 tunneling (ferromagnetic case), the tunneling
Hamiltonian can analogously be written as
\begin{eqnarray}
H_{T} &=&
-t^{\prime\prime}_{1} d^{\dagger} a^{\dagger}_{\downarrow}
-t^{\prime\prime}_{2}\, (d^{\dagger}a^{\dagger}_{\uparrow}
          + \gamma_{z}a^{\dagger}_{\downarrow})
\nonumber\\&&
{}+t^{\prime\prime}_{3}\, (-\gamma_{z} a^{\dagger}_{\uparrow}
           + da^{\dagger}_{\downarrow})
+ t^{\prime\prime}_{4} d a^{\dagger}_{\uparrow}
+\text{H.c.},
\label{eq:Hd32}
\end{eqnarray}
where $t''_1\equiv t^{3/2,-3/2}$, $t''_2\equiv t^{3/2,-1/2}/\sqrt{3}$,
$t''_3\equiv t^{3/2,1/2}/\sqrt{3}$, $t''_4\equiv t^{3/2,3/2}$.
The tunneling amplitudes $t^{3/2,m}$ form a spin-$3/2$ spinor.
By an appropriate choice of spin basis we can again set one of the
$t^{3/2,m}$ (and thus the corresponding $t''_n$) to zero.
However, no matter which tunneling amplitude is set to zero, both
spin channels remain coupled through the $d$ fermion and thus, directly
or indirectly, to the Majorana fermion $\gamma_z$. Under a bias, the current
is non-zero for all electron-spin states.
Therefore, in general the current for spin-3/2 tunneling can only be
partially spin polarized.


\textit{Summary}.---The presence of the Kondo interaction between the
electrons in the normal lead and Majorana fermions at the ends
of a TS wire results in a strong
renormalization of the tunneling processes through the interface. The tunneling
amplitudes can be classified according to the irreducible representations of
the SU(2) group. The amplitudes belonging to different
representations obey different scaling laws. Depending on the sign of
the interaction, one component is enhanced, while the other is
suppressed so that only one type of tunneling survives.
Ferromagnetic interaction favors a spin-3/2
tunneling with parallel electron spin and Majorana pseudospin,
whereas antiferromagnetic coupling enhances spin-1/2 tunneling with
opposite spin and pseudospin. The temperature dependence and spin polarization
of the current through the N--TS interface reflects the presence of
multiple Majorana states and the type of interaction, and therefore can be used
as a tool for the search of a topological system with multiple
edge states and for the determination of their interaction type.

\begin{acknowledgments}
Financial support by the Deutsche Forschungsgemeinschaft through Research
Training Group GRK 1621 is gratefully acknowledged. We would like to
thank M. Vojta for useful discussions.
\end{acknowledgments}

\bibliographystyle{apsrev4-1}
\bibliography{mlcoupling}

\end{document}